\documentclass[article, notitlepage,superscriptaddress, longbibliography]{revtex4-2}
\usepackage{natbib}
\usepackage{graphicx}
\usepackage{amssymb}
\usepackage{amsmath}
\usepackage{ifthen}
\usepackage{braket}
\usepackage{xcolor}
\usepackage{bm}
\usepackage{bbm}
\usepackage{bbm}
\usepackage{maybemath}
\usepackage{comment}
\usepackage{siunitx}
\usepackage{float}
\usepackage{dcolumn}
\newcolumntype{d}[1]{D{.}{.}{#1}}
\usepackage{subfigure}

\newcommand{\dd}{\mathrm{d}}
\newcommand{\ii}{\mathrm{i}}
\newcommand{\ee}{\mathrm{e}}

\def\dd{{\mathrm{d}}}
\def\ii{{\mathrm{i}}}
\def\ee{{\mathrm{e}}}

\def\bfT{{\bf{T}}}
\def\rmT{{\mathrm{T}}}
\def\calT{{\mathcal{T}}}

\newcommand{\hate}{\hat{\mathrm{e}}}

\definecolor{garrosgreen}{rgb}{0.1, 0.4, 0.1}
\definecolor{dartmouthgreen}{rgb}{0.05, 0.5, 0.06}
\definecolor{duelferred}{rgb}{0.7, 0.2, 0.1}
\definecolor{cambridgeblue}{rgb}{0.1, 0.3, 1.0}
\definecolor{oxfordblue}{rgb}{0.05, 0.2, 0.7}

\newcommand{\addrFSU}{Department of Chemistry, Physics and Materials Science,
Fayetteville State University, Fayetteville, North Carolina 28301, USA}

\newcommand{\addrRolla}{Department of Physics and LAMOR,
Missouri University of Science and Technology,
Rolla, Missouri 65409, USA}

\newcommand{\addrDebrecen}{Hungarian Academy Institute
for Nuclear Physics (ATOMKI), Debrecen, Hungary}

\begin{document}

\title{Quantum Electrodynamics of Dicke States:\\
Resonant One-Photon Exchange Energy and Entangled Decay Rate}

\author{U. D. Jentschura}
\affiliation{\addrRolla}
\affiliation{\addrDebrecen}

\author{C. M. Adhikari}
\affiliation{\addrFSU}

\begin{abstract}
We calculate the one-photon
exchange contribution to the interatomic interaction potential
between electrically neutral, identical atoms, 
one of which is assumed to be in an excited state,
by matching the scattering matrix ($S$ matrix) element with the 
effective Hamiltonian. This approach allows us to use covariant 
perturbation theory, where the two possible time orderings
of emission and absorption are summarized in a single Feynman amplitude. 
Our results encompass the full retardation correction to the
one-photon exchange van-der-Waals potential.
We employ the temporal gauge for the virtual photon
propagator. Based on the Feynman prescription,
we obtain the imaginary part of the interaction energy,
which leads to an interaction-induced 
correction to the decay rate. Our results
lead to precise formulas for the distance-dependent enhancement
and suppression of the decay rates of entangled 
superradiant and subradiant
Dicke state, as a function of the interatomic distance.
We apply the result to an example calculation involving two hydrogen
atoms, one of which is in the ground state,
the other in an excited $P$ state.
\end{abstract}

\maketitle

%
% Introduction
%
\section{Introduction}
\label{sec1}

Normally, one assumes that 
interatomic long-range interactions 
(van der Waals interactions) are associated 
with fluctuating dipole moments of
the two interacting atoms, which, in turn,
are due to quantum fluctuations 
of the electron positions in both atoms.
In the dipole approximation,
the interaction Hamiltonian is 
equal to the scalar product
of the dipole operator and the 
second-quantized electric-field 
operator. If both atoms are in an initial
ground state (which is generally 
spherically symmetric), then both of them
undergo a virtual dipole transition 
to an excited state upon emission 
(atom $A$) and absorption 
(atom $B$) of a virtual photon.
The process is complete upon the exchange
of a second virtual photon,
with the order of emission 
(atom $B$) and absorption 
(atom $A$) reversed. The final state
has both atoms in the ground state,
without any photons present
(no excitations of the photon field).
The energy of the final state 
is equal to that of the initial state,
but energy conservation does not hold for
the virtual transitions.

Because each emission and absorption of 
a virtual photon involves a single power
of the interaction Hamiltonian,
the exchange of two photons (two emissions
and two absorptions) is a process of 
fourth-order perturbation theory.
The corresponding interaction energy 
can be calculated in both time-ordered
perturbation theory (see Ref.~\cite{CrTh1984} 
and~Chap.~5 of Ref.~\cite{JeAd2022book})
and by matching the scattering amplitude 
to the effective Hamiltonian
(see Ref.~\cite{JeDe2017vdw0} and Chap.~12 
of Ref.~\cite{JeAd2022book}).
The above situation pertains
to two interacting ground-state atoms,
and the resulting expressions contain 
the polarizabilities of both involved atoms.
The result (for both identical as well as 
non-identical {\em ground-state} atoms)
smoothly interpolates between the non-retarded
short-range $1/R^6$ so-called van der Waals limit
and the fully retarded $1/R^7$ long-range limit
of the interaction.
The process involves the exchange 
of two virtual photons 
(see Ref.~\cite{CrTh1984}
and~Chap.~5 of Ref.~\cite{JeAd2022book}).

However, if we consider two identical atoms,
one of which is in an excited state, there can be energetically
degenerate states of the two-atom system
which are connected with the initial 
state of the process by the exchange 
of a single, not two, virtual photons. 
In the quantum-field theoretical 
picture, this situation corresponds to a nonvanishing 
transition matrix elements of the interaction
Hamiltonian already in the second-order perturbation 
theory of the second-quantized Hamiltonian.
For example, if one of the atoms (atom $A$) is a
hydrogen atom in a $1S$ state,
and the other (atom $B$) is a hydrogen atom in a $2P$ state,
then there can be a one-photon exchange process
which connects the initial state 
to a state where atom $A$ is in the $2P$,
and atom $B$ is in the $1S$ state.
The final state is energetically fully degenerate
with the initial state of the 
process. This implies that the eigenstates
of the combined Hamiltonian of the individual
atoms, and of the interaction, are coherent
superpositions of product states of atoms $A$ 
and $B$ in the $| (1S)_A (2P)_B \rangle$, 
$| (2P)_A (1S)_B \rangle$ states. (An illustrative 
discussion can also be found near the beginning of Chap.~7 of
Ref.~\cite{CrTh1984}.)

In the nonretarded approximation, this process
has been treated in Refs.~\cite{JoEtAl2002,JeAd2017atoms,AdEtAl2017vdWi,JeEtAl2017vdWii}.
One observes that the treatment in these 
references relies on the use of the 
van-der-Waals interaction Hamiltonian
\begin{equation}
\label{HvdW}
H_{\rm vdW} =
\left( \delta_{ij}-3 \hat R_i \, \hat R_j  \right)
\frac{ d_{Ai} \, d_{Bj} }{4 \pi \epsilon_0 R^3} \,
\end{equation}
which involves the product of the 
dipole operators of both atoms.
Here, $\vec R$ is the interatomic separation 
vector, $\hat R = \vec R/R$ is its unit vector,
$\delta_{ij}$ is the Kronecker symbol,
and $\vec d_{A}$ and $\vec d_B$ are the 
dipole operators for the two atoms.
It leads to a nonvanishing matrix element
in the energetically 
degenerate system in the first order of perturbation theory.
One might ask how can this be understood, when we just said that
one-photon exchange is a second-order perturbative
process. The answer is that the van der Waals
Hamiltonian is derived without field quantization, i.e.,
by simply expanding the electrostatic (instantaneous)
Coulomb interactions of the constituent electrons and nuclei
in the two atoms~\cite{JoEtAl2002,JeAd2017atoms,AdEtAl2017vdWi,JeEtAl2017vdWii},
in powers of the distances of the
electrons and nuclei. This expansion
does not use field quantization.
A single nonretarded Coulomb interaction
is proportional to $e^2$, where $e$ is the electron charge,
and is thus of second order in the quantum-field theoretical
picture, where each photon emission or absorption vertex
is considered to add an order of perturbation theory.
In the second-quantized picture, one uses
{\em temporal gauge} 
[see Eq.~(9.133) of Ref.~\cite{JeAd2022book}]
and describes the same process
in second-order perturbation theory, using two 
second-quantized interaction Hamiltonians
which are each proportional to the scalar product of 
dipole operators and second-quantized electric field.
Because the dipole operators are proportional to $e$,
the resulting interaction also is proportional to $e^2$.
The timelike component of the photon propagator 
vanishes in temporal gauge, and it is therefore
ideally suited to treat the retarded form of the 
van der Waals interaction.

Implicitly, the temporal gauge 
actually is used in the derivation outlined 
in Chap.~7 of Ref.~\cite{CrTh1984},
where the interaction with the radiation field
is formulated exclusively in terms of the 
dipole coupling term with the electric field.
A decisive difference 
to the derivation outlined here is that,
in our covariant approach, the consideration 
of two different time orderings of the 
photon emission and absorption 
by the two atoms involved in the interaction.
Hence, by using the technique 
of matching the effective Hamiltonian with the 
scattering matrix element, we can unify both time orderings 
into one single Feynman diagram.
The most interesting consequence of the use of the 
Feynman prescription is the emergence of an 
imaginary part of the one-photon exchange interaction,
which leads to a modification
of the decay rate.

The modification of the decay rate is especially 
interesting for {\color{black} two-atom} Dicke states~\cite{Di1953},
{\color{black} otherwise known as Bell states~\cite{wikibellstate},}
which constitute entangled states of the two-atoms system.
If we denote the ground state as $| \psi_g \rangle$ 
and the excited state as $| \psi_e \rangle$, then the 
Dicke states are
\begin{align}
\label{dicke}
| \Psi_\pm \rangle 
=& \; \frac{1}{\sqrt{2}} \,
\left[ | (\psi_e)_A (\psi_g)_B \rangle
\pm | (\psi_g)_A (\psi_e)_B \rangle \right]
\\
=& \; \frac{1}{\sqrt{2}} \,
\left[ | \psi_e \, \psi_g \rangle
\pm | \psi_g \, \psi_e \rangle \right] \,.
\end{align}
Let us briefly discuss the entanglement.
In the basis of states
\begin{equation}
| \psi_1 \rangle = | \psi_g \rangle  
\,, \quad
| \psi_2 \rangle = | \psi_e \rangle 
\end{equation}
one can form the following two-particle states,
\begin{equation}
| \Phi_1 \rangle = | \psi_g \, \psi_g \rangle 
\,, \quad
| \Phi_2 \rangle = | \psi_e \, \psi_g \rangle
\,, \quad
| \Phi_3 \rangle = | \psi_g \, \psi_e \rangle
\,, \quad
| \Phi_4 \rangle = | \psi_e \, \psi_e \rangle \,.
\end{equation}
Dicke states have the form
\begin{align}
| \Psi_\pm \rangle
=& \; 
\frac{1}{\sqrt{2}} \, | \Phi_2 \rangle
\pm \frac{1}{\sqrt{2}} \, | \Phi_3 \rangle
= \sum_{i=1}^2 \sum_{j=1}^2 
c^\pm_{ij} \; | \psi_i \psi_j \rangle \,,
\qquad
c^\pm_{12} = \frac{1}{\sqrt{2}} \,,
\qquad
c^\pm_{21} = \pm c^\pm_{12} \,,
\end{align}
while all $c^\pm_{ij}$ other than $c^\pm_{12}$
and $c^\pm_{21}$ vanish. It is crucial to 
observe that these $c_{ij}$ cannot be written in the form
$c_{ij} = a_i \, b_j$, and the Dicke states 
are thus entangled.

As a clarifying remark, we do not consider transient
phenomena connected to the 
{\color{black}
excitation process~\cite{MiKn1974,BiEtAl1990,BiEtAl1990,PlWiHiLu1990,He1994,MiJaFe1995,%
BeDu1997,PoTh1997,Be2015},
and [see also Eqs.~(5.10),~(5.11) and (5.16) 
of Ref.~\cite{PoTh1997}]
and consider the entangled Dicke states as the 
basis of our discussions.}
This paper studies the properties, not the preparation,
of the entangled Dicke states.
{\color{black} Following p.~200 of Ref.~\cite{Sa2010},
we remark that the concept of an intermolecular interaction 
energy for the situation in which the initial state of A or B corresponds
to an excited state holds as long as the excited state or states in question are
sufficiently long lived relative to the time taken for the photon to propagate
between the two sites. The preparation of Dicke states by 
carefully engineered light pulses 
has been discussed in Eqs.~(4.21) and Eqs.~(7.1), (7.2) and~(7.3) 
of Ref.~\cite{Ri2005}.
Further considerations on suitable preparation algorithms
have been reported in Refs.~\cite{RiEtAl2008,HuChRoWi2009,BaEi2020}.
We note that the entangled denoted here as 
$| \Psi_+ \rangle$ is known as the Bell state 
$| \Psi^+ \rangle$ in the literature on quantum 
computation~\cite{Ri2005,wikibellstate}.}

{\color{black} From a historical perspective, it is 
interesting to remark that the 
possibility of resonant energy transfer between 
excited and ground states of atomic and molecular 
systems via the exchange of resonant virtual photons
has been recognized in the early days of quantum
mechanics~\cite{CaFr1922,Pe1927,Pe1932,Fe1932},
and summarized in reviews~\cite{Fo1946,Fo1948,JoBr2019}.
Thus, considerable effort has been invested 
into the calculation of the retardation 
corrections to the interaction potential 
given in Eq.~\eqref{HvdW}
(see Refs.~\cite{MLPo1964,AnSh1987,An1989,DaJeBrAn2003,JeDaAn2004,%
Sa2015,JeDaAn2016,Sa2018,JoBr2019}).}

The {\color{black} three} advantages of the second-quantized picture are
that {\em (i)} it becomes possible to study the effect of 
retardation, i.e., the effect of the 
finite speed of light is the propagation of the 
interaction from atom $A$ to atom $B$,
and {\em (ii)} it is possible to obtain precise 
formulas for the distance-dependent modification of 
the decay rate of Dicke states.
{\color{black} Furthermore, {\em (iii)} the 
field-theoretical formalism employed here 
allows us to consistently identify the position 
of the poles of the propagator denominators,
in view of a consistent application
of the Feynman prescription~\cite{JeAd2022book} which is implicit in the 
matching procedure employed here.
Our goal is to obtain the interaction potential between
atoms, on account of retardation, 
as a function of the variable $\omega R/c$,
where $\omega$ is an angular frequency of a photon,
$R$ is the interatomic distance,
and $c$ is the speed of light.

Our calculation pertains to the exchange
of one, not two, photons. 
In the contrasting case of the two-photon interaction,
$\omega$ is the modulus of the 
frequency of either of the two virtual photons;
the frequencies add up to zero in view of 
energy conservation in the initial and final 
states~\cite{JeDe2017vdw0}.
Thus, the relevant diagrams for the one-photon exchange
are {\em not} those given in Fig.~5.1 of Ref.~\cite{JeAd2022book},
{\em not} those given in Fig.~7.5 of Ref.~\cite{CrTh1984},
{\em not} those given in Fig.~1 of Ref.~\cite{AlSa2006},
but those given in Fig.~1 of Ref.~\cite{DaJeBrAn2003},
and Fig.~2 of Ref.~\cite{JoBr2019}
(and in Fig.~\ref{fig1} here).}
Natural units with $\hbar = c = \epsilon_0$ 
are used in the following unless stated otherwise.

%
% $\mathbebm{S}$--Matrix and Matching with Effective Interaction
%
\section{$\maybebm{S}$--Matrix and Effective Hamiltonian}
\label{sec2}

\begin{figure}
\begin{center}
\begin{minipage}{0.8\linewidth}
\begin{center}
\includegraphics[width=0.8\linewidth]{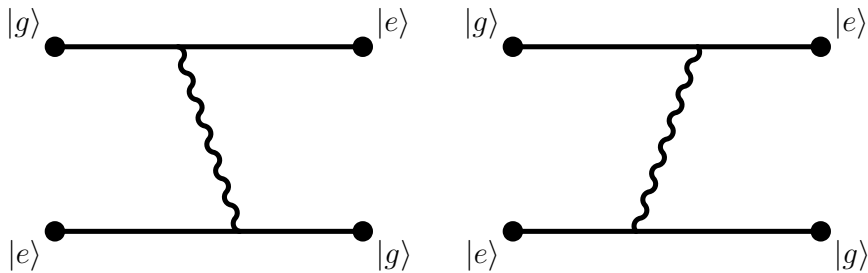}
\caption{\label{fig1} Diagrams are given for the exchange 
of a virtual photon of angular frequency $\omega_0$ 
between two identical atoms.
The two different time orderings are written
out explicitly for illustration, 
even if, in the language of Feynman diagrams
of quantum electrodynamics~\cite{JeAd2022book}, the diagrams
are considered to be identical. 
{\color{black} The ground state is denoted as $| g \rangle$,
the excited state as $| e \rangle$.}}
\end{center}
\end{minipage}
\end{center}
\end{figure}

It is useful to recall the principle of 
matching the $\maybebm{S}$ matrix 
with the effective Hamiltonian.
We consider two identical atoms $A$ and $B$ in the initial states 
$| g \rangle$ and $| e \rangle$ (ground and excited),
\begin{equation}
\label{scattering1}
\psi_{A}(\vec r_A) = \psi_g(\vec x_A - \vec R_A) \,,
\qquad
E_A = E_g \,,
\qquad
\psi_B(\vec r_B) = \psi_e(\vec x_B - \vec R_B) \,,
\qquad
E_B = E_e \,,
\end{equation}
which scatter into the final states
\begin{equation}
\label{scattering2}
\psi_A'(\vec r_A) = \psi_e(\vec x_A - \vec R_A) \,,
\qquad
E'_A = E_e \,,
\qquad
\psi_B'(\vec r_B) = \psi_g(\vec x_B - \vec R_B) \,,
\qquad
E'_B = E_g \,,
\end{equation}
under the action of a potential $U$.
The electron coordinates are $\vec x_A$ and $\vec x_B$,
and the coordinates of the nuclei are $\vec R_A$ and $\vec R_B$.
The energies of the states $| g \rangle$ and $| e \rangle$
are assumed to differ by $\omega_0$, where
\begin{equation}
\label{omega0}
\omega_0 = | E_A - E_B | = | E'_A - E'_B | 
= | E_e - E_g | \,.
\end{equation}
The final state of the process has the 
energy of the two atomic states 
interchanged and therefore
the same energy as the initial state.
The corresponding Feynman diagrams are 
given in Fig.~\ref{fig1}.

For the matching, we consider the action of the potential 
$U(\vec r_A, \vec r_B, \vec R)$, 
where $\vec R = R_A - \vec R_B$ is the 
interatomic distance.
The corresponding (first-order) $S$-matrix element 
reads as follows,
\begin{align}
S_{fi} =& \; -\ii \,
\int \dd^3 r_A \int \dd^3 r_B \;
\psi_A'^*(\vec r_A) \, \psi_B'^*(\vec r_B) \,
U(\vec r_A, \vec r_B, \vec R) \,
\psi_A(\vec r_A) \, \psi_B(\vec r_B) \,
\nonumber\\
& \; \times
\int \dd t \, \ee^{ -\ii \, (E_A + E_B - E'_A - E'_B) \, t }
\nonumber\\
=& \; -\ii \, T \, 
\int \dd^3 r_A \int \dd^3 r_B \;
\psi_A'^*(\vec r_A) \, \psi_B'^*(\vec r_B) \,
U(\vec r_A, \vec r_B, \vec R) \,
\psi_A(\vec r_A) \, \psi_B(\vec r_B) \,.
\end{align}
We have assumed energy conservation
($E_A + E_B = E'_A + E'_B$) and 
denoted the (long) time interval over which the 
transition from initial to final state occurs, 
as $\int \dd t = T$. In our calculations,
we shall obtain the (manifestly) non-forward 
scattering amplitude (the final states differ from the 
initial states) in the functional form 
\begin{align}
S_{fi} =& \; -\ii \,
\int \dd^3 r_A \int \dd^3 r_B \;
{\psi'_{A}}^*(\vec r_A) \, {\psi'_B}^*(\vec r_B) \,
S(\vec r_A, \vec r_B, \vec R)  \,
\psi_A(\vec r_A) \, \psi_B(\vec r_B) 
\nonumber\\
=& \; -\ii \,
\left< {\psi'_{A}} \, {\psi'_B} \left| 
S \right| \psi_A \, \psi_B \right> \,.
\end{align}
The matching relation is thus 
\begin{equation}
\label{match}
\langle S(\vec r_A, \vec r_B, \vec R) \rangle =
T \, U( \vec r_A, \vec r_B, \vec R ) \,,
\qquad
\vec R = \vec R_A - \vec R_B \,.
\end{equation}
The use of the notation $\langle S(\vec r_A, \vec r_B, \vec R) \rangle$
indicates that we obtain the matching in the integrand 
of the scattering matrix element.
The final integration occurs over the wave functions 
of the initial and final states of the two-atom system.

%
% Interaction Hamiltonian
%
\section{Calculation of the Effective Hamiltonian}
\label{sec3}

In time-dependent quantum 
electrodynamic (QED) perturbation theory,
the interaction is formulated in the 
interaction picture~\cite{ItZu1980,MoPlSo1998,JeAd2022book}.
The second-quantized operators in the interaction
Hamiltonian have a time dependence 
which is generated by the action of the 
free Hamiltonian (see Chap.~3 of Ref.~\cite{JeAd2022book}).
The designation of an interaction Hamiltonian
being formulated in the interaction picture 
is not redundant~(see Chap.~3 of Ref.~\cite{JeAd2022book}).
The interaction Hamiltonian is
\begin{equation}
V(t) = 
-\vec E(\vec R_A, t) \cdot \vec d_A(t)
-\vec E(\vec R_B, t) \cdot \vec d_B(t) \,.
\end{equation}
Here, $\vec d_k = e \, \vec r_k = e \, (\vec x_k - \vec R_k)$
with $k=A,B$ is the dipole operator for atom $k$ 
(for atoms with more than one electron, one has to sum over all the electrons).
The $\vec R_A$ and $\vec R_B$
are the positions of the atomic nuclei,
while the $\vec x_A$ and $\vec x_B$ denote the 
electron coordinates.
The second-order contribution to the $S$-matrix is
\begin{align}
\label{S2start}
\langle \phi', 0 | S^{(2)} | \phi, 0
\rangle =& \; \frac{(-\ii)^2}{2!}
\int \dd t_1 \int \dd t_2 
\; \left< \phi', 0 \left| {\bfT}[ V(t_1) V(t_2) 
{\color{black} ] } \right| \phi, 0 \right>  \,.
\end{align}
Here, ${\bfT}$ denotes the ordering of 
all operators, pertaining to both atomic 
dipoles and electric fields.
We denote by $| 0 \rangle$ the ``vacuum''
of the electromagnetic field.
In the vacuum, there are no photons in the radiation field.
Let us also denote the initial state as
$| \phi \rangle = | \psi_A = \psi_g, \, \psi_B = \psi_e  \rangle$ and
the final two-atom state as
$| \phi' \rangle = | \psi'_A = \psi_e, \, \psi'_B = \psi'_g  \rangle$,
with the conditions given in Eqs.~\eqref{scattering1}
and~\eqref{scattering2}. Then, 
\begin{align}
\left< \phi', 0 \left| \bfT[ V(t_1) V(t_2) 
{\color{black} ] } \right| \phi, 0 \right> =& \;
\left< \phi', 0 \left| \bfT
\left[ 
\left( -\vec E(\vec R_A, t_1) \cdot \vec d_A(t_1)
-\vec E(\vec R_B, t_1) \cdot \vec d_B(t_1) \right) \,
\right. \right. \right. 
\nonumber\\[2ex]
& \; 
\left. \left.  \left. 
\times \left( -\vec E(\vec R_A, t_2) \cdot \vec d_A(t_2)
- \vec E(\vec R_B, t_2) \cdot \vec d_B(t_2) \right) 
\right] 
\right| \phi, 0 \right> 
\nonumber\\[2ex]
\sim & \;
\left< \phi', 0 \left| \bfT
\left[ \left( \vec E(\vec R_A, t_1) \cdot \vec d_A(t_1) \right) \,
\left( \vec E(\vec R_B, t_2) \cdot \vec d_B(t_2) \right) \right]
\right| \phi, 0 \right>
\nonumber\\[2ex]
& \; + \left< \phi', 0 \left| \bfT
\left[ \left( \vec E(\vec R_B, t_1) \cdot \vec d_B(t_1) \right) \,
\left( \vec E(\vec R_A, t_2) \cdot \vec d_A(t_2) \right) \right]
\right| \phi, 0 \right> \,,
\end{align}
where by $\sim$ we denote the omission of operators 
which pertain to the one-loop self energy of the two atoms,
and keep only the terms relevant for the interaction energy.
The time ordering of the electric field operators
and atomic dipole operators leads to 
\begin{align}
\label{S2_expr1}
S^{(2)} =& \; -\frac{1}{2} \int \dd t_1 \int \dd t_2 \,
\left[ \left< 0 \left|
\calT\left[ E^i(\vec R_A, t_1) \, E^j(\vec R_B, t_2) \right]
\right| 0 \right> \,
\left< \psi_e \, \psi_g \left|
\rmT \; d_A^i( t_1) \, d_B^j( t_2) 
\right| \psi_g \, \psi_e \right> 
\right.
\nonumber\\[0.1133ex]
& \; -\frac{1}{2} \int \dd t_1 \int \dd t_2 \,
\left[
\left< 0 \left|
\calT\left[ E^i(\vec R_B, t_1) \, E^j(\vec R_A, t_2) \right]
\right| 0 \right> \,
\left< \psi_e \, \psi_g \left|
\rmT \; d_B^i( t_1) \, d_A^j( t_2)
\right| \psi_g \, \psi_e \right> 
\right. \,.
\end{align}
Here, $\rmT$ is the time ordering operator 
for the atomic dipole moments, while the 
$\calT$ operator time-orders the electric field operators. 
Also, $i,j = 1,2,3$ denote the 
Cartesian components.
We use a relativistic notation where the superscript
denotes the Cartesian component.

We now recall a few known results from 
Ref.~\cite{JeDe2017vdw0} and Chap.~12 of~\cite{JeAd2022book}.
The time-ordered product of electric-field operators
can be evaluated as follows,
\begin{equation}
\left< 0 \left|
\calT\left[ E^i(\vec R_A, t_1) \, E^j(\vec R_B, t_2) \right]
\right| 0 \right> = 
\ii \, \int \frac{\dd \omega}{2 \pi} \,
\omega^2 \, D_F^{ij}(\omega, \vec r) \, \ee^{-\ii \omega (t_1 - t_2)} \,.
\end{equation}
Here,
\begin{equation}
D_F^{ij}(\omega, \vec R) =
- \left( \delta^{ij} + \frac{\nabla^i \, \nabla^j}{\omega^2} \right) \,
\frac{\ee^{\ii |\omega| R}}{4 \pi R}  \,,
\qquad |\omega| = \sqrt{\omega^2 + \ii \epsilon} \,,
\end{equation}
is the photon propagator in the mixed frequency-position 
representation, in the temporal gauge 
[in the conventions outlined in Eq.~(9.133) of Ref.~\cite{JeAd2022book}].
In the temporal gauge, the timelike component of the 
photon propagator vanishes, and one has 
$D_{00}(\omega, \vec R) = 0$.
We use the photon propagator in the convention $\ii D_F^{\mu\nu}(x - x') =
\left< 0 \left| \calT A^\mu(x) \, A^\nu(x') \right| 0 \right>$,
with $\mu,\nu=0,1,2,3$, where $A^\mu(x)$ is the 
quantized four-vector potential propagator. 
The result for the spatial components of the 
photon propagator {\color{black} is found 
according to Refs.~\cite{CrTh1989,JeDe2017vdw0}},
\begin{equation}
\label{Dij}
D_F^{ij}(\omega, \vec R) =
- \left[ \alpha^{ij} + \beta^{ij} \,
\left( \frac{\ii}{|\omega| R} -
\frac{1}{\omega^2 \, R^2} \right) \right] \,
\frac{\ee^{\ii |\omega| R} }{4 \pi R} \,,
\quad
\alpha^{ij} = \delta^{ij} - {\hat R}^i \, {\hat R}^j \,,
\quad
\beta^{ij} = \delta^{ij} - 3 {\hat R}^i \, {\hat R}^j \,,
\end{equation}
where $|\omega| = \sqrt{\omega^2 + \ii \epsilon}$,
and the branch cut of the square root is taken
along the positive real axis~\cite{Mo1974a,Mo1974b}.
Now, let us proceed to evaluate the
time-ordered product of dipole operators,
\begin{equation}
X^{ij}(t_1 - t_2) 
= \left< \psi_e \, \psi_g \left| 
\rmT \; d_A^i(t_1) \, d_A^j(t_2) \right| \psi_g \, \psi_e \right> 
= \left< \psi_e \, \psi_g \left| 
\rmT \; d_A^i(t_1 - t_2) \, d_A^j(0) \right| \psi_g \, \psi_e \right> \,.
\end{equation}
Introducing the Fourier transform $X^{ij}(\omega)$,
we can write
$X^{ij}(t_1 - t_2) = \int
\frac{\dd \omega}{2 \pi} \, \ee^{-\ii \, \omega \, (t_1 - t_2)}  \,
X^{ij}(\omega)$.
The Fourier transform of the 
time-ordered product of dipole operators can be evaluated 
as follows,
\begin{align}
\label{alphaFEYNMAN}
X^{ij}(\omega) =& \; \int_{-\infty}^\infty \dd t \, 
\ee^{\ii \omega t} \, X^{ij}(t) 
= \int_{-\infty}^\infty \dd t \, \ee^{\ii \omega t} \, 
\left< \psi_e \, \psi_g \left| 
\rmT \; d_A^i(t) \, d_B^j(0) \right| \psi_g \, \psi_e \right>
\nonumber\\[0.1133ex]
=& \; \int_{0}^\infty \dd t \, \ee^{\ii \omega t} \,
\left< \psi_e \, \psi_g 
\left| d_A^i(t) \, d_B^j(0) \right| \psi_g \, \psi_e \right>
+ \int_{-\infty}^0 \dd t \, \ee^{\ii \omega t} \,
\left< \psi_e \, \psi_g \left| d_B^j(0) \, d_A^i(t) 
\right| \psi_g \, \psi_e \right>
\nonumber\\[0.1133ex]
=& \; 
\left( \int_{0}^\infty 
+ \int_{-\infty}^0 \right)
\dd t \, \ee^{\ii \omega t} \,
\left< \psi_e \left| d_A^i(t) \right| \psi_g \right> \, 
\left< \psi_g \left| d_B^j(0) \right| \psi_e \right> \,.
\end{align}
We have used the fact that the atoms are identical 
and undergo transitions $| g \rangle \to |e \rangle$
and $| e \rangle \to |g \rangle$, respectively.
We can thus add the two integration domains and 
conclude that 
\begin{align}
X^{ij}(\omega) 
=& \; \int_{-\infty}^\infty \dd t \, \ee^{\ii \omega t} \,
\left< \psi_e \left| d_A^i(t) \right| \psi_g \right> \,
\left< \psi_g \left| d_B^j(0) \right| \psi_e \right>
\nonumber\\[0.1133ex]
=& \; \int_{-\infty}^\infty \dd t \, \ee^{\ii \omega t} \,
\ee^{\ii (E_2 - E_1) \, t} \,
\left< \psi_e \left| d_A^i(0) \right| \psi_g \right> \,
\left< \psi_g \left| d_B^j(0) \right| \psi_e \right>
\nonumber\\[0.1133ex]
=& \; 2 \pi \delta(E_2 - E_1 + \omega) \,
\left< \psi_e \left| d_A^i(0) \right| \psi_g \right> \,
\left< \psi_g \left| d_B^j(0) \right| \psi_e \right> \,.
\end{align}
The two terms in Eq.~\eqref{S2_expr1} 
yield equivalent contributions, 
and we obtain, with the help of the results obtained previously
for the time-ordered products of the 
electric-field and atomic dipole operators,
\begin{align}
S^{(2)} =& \;
\frac{1}{2} \int \dd t_1 \int \dd t_2 \,
\left< 0 \left|
\calT\left[ E^i(\vec R_A, t_1) \, E^j(\vec R_B, t_2) \right]
\right| 0 \right> \,
\left< \psi_e \, \psi_g \left|
\rmT \; d_A^i( t_1) \, d_B^j( t_2)
\right| \psi_g \, \psi_e \right>
\nonumber\\[0.1133ex]
=& \;
\int \dd t_1 \int \dd t_2 \,
\left[
-\ii \, \int \frac{\dd \omega}{2 \pi} \,
\omega^2 \, D_F^{ij}(\omega, \vec R) \, \ee^{-\ii \omega (t_1 - t_2)} 
\right]
\ee^{-\ii \, (E_1 - E_2)\, (t_1 - t_2)}  \,
\nonumber\\[0.1133ex]
& \; \times
\left< \psi_e \left| d_A^i \right| \psi_g \right> \,
\left< \psi_g \left| d_B^j \right| \psi_e \right> 
\nonumber\\[0.1133ex]
=& \;
-\ii \int \dd (t_1 - t_2) \int \dd t_2 
\int \frac{\dd \omega}{2 \pi} 
\omega^2 \, D_F^{ij}(\omega, \vec R)  
\ee^{-\ii (\omega + E_1 - E_2) \, (t_1 - t_2)} 
\left< \psi_e \left| d_A^i \right| \psi_g \right> 
\left< \psi_g \left| d_B^j \right| \psi_e \right>
\nonumber\\[0.1133ex]
% %
% =& \;
% -\ii \, T \, \int \frac{\dd \omega}{2 \pi} \,
% \omega^2 \, D_F^{ij}(\omega, \vec R) \,
% 2 \pi \, \delta(\omega + E_1 - E_2) \, 
% \left< \psi_e \left| d_A^i \right| \psi_g \right> \,
% \left< \psi_g \left| d_B^j \right| \psi_e \right>
% \nonumber\\[0.1133ex]
%
=& \;
-\ii \, T \, 
(E_2 - E_1)^2 \, D_F^{ij}( | E_2 - E_1 |, \vec R) \,
\left< \psi_e \, \psi_g \left| d_A^i \, d_B^j \right| 
\psi_g \, \psi_e \right> \,,
\end{align}
where $d_A^i \equiv d_A^i(0)$ and $d_{Bj} \equiv d_{Bj}(0)$.
Based on the matching relation~\eqref{match}
and remembering Eq.~\eqref{omega0},
we read off the interaction potential,
\begin{align}
\label{U}
U(\vec r_A, \vec r_B, \vec R) = & \;
(E_2 - E_1)^2 \, D^{ij}( | E_2 - E_1 |, \vec R) \,
d_A^i \, d_B^j
\nonumber\\[0.1133ex]
=& \; -\omega_0^2 \, 
\left[ \alpha^{ij} + \beta^{ij} \,
\left( \frac{\ii}{\omega_0 \, R} -
\frac{1}{\omega_0^2 \, R^2} \right) \right] \,
\frac{\ee^{\ii \omega_0 \, R} }{4 \pi R} \,
d_A^i \, d_B^j
\nonumber\\[0.1133ex]
=& \; 
\left[ \left( \delta^{ij} - 3 \frac{R^i \, R^j}{R^2} \right) \,
\left( 1 - \ii \omega_0 R \right) 
- \left( \delta^{ij} - \frac{R^i \, R^j}{R^2} \right) 
\omega_0^2 \, R^2 \right] \,
\frac{ \ee^{\ii \omega_0 \, R} }{4 \pi R^3} \,
d_A^i \, d_B^j \,.
\end{align}
{\color{black} This result confirms 
results previously obtained in 
Eqs.~(13), (14), (40), (41) and (42) of Ref.~\cite{AnSh1987},
Eqs.~(12), (13), (14) of Ref.~\cite{DaJeBrAn2003},
Eq.~(2.23) of Ref.~\cite{JeDaAn2004},
Eqs.~(4) and (5) of Ref.~\cite{Sa2015},
Eq.~(2.1) of Ref.~\cite{JeDaAn2016},
Eqs.~(3.5), (3.6) and (3.7) of Ref.~\cite{An1989},
Eq.~(14) of Ref.~\cite{Sa2018},
and Eq.~(5) of Ref.~\cite{JoBr2019}.
One important advantage of the method of derivation
employed here is that the imaginary part 
follows directly from the Feynman prescription
for the propagator denominators. 
In other approaches, additional considerations are
required to fix the location of the poles 
of the propagators in the complex plane
(for an illustrative discussion on this
point, see Ref.~\cite{JoBr2019}).}
This potential can be separated into a real and an 
imaginary part,
\begin{subequations}
\begin{align}
\label{ReU}
{\rm Re} [U(\vec r_A, \vec r_B, \vec R)] = & \;
\left[ 
\beta^{ij} \, \left( 
\cos( \omega_0 \, R ) + \omega_0 R \, \sin( \omega_0 \, R ) 
\right) 
- \alpha^{ij} \, \omega_0^2 \, R^2 \, \cos( \omega_0 \, R ) 
\right] 
\, \frac{d_A^i \, d_B^j}{4 \pi R^3} \,,
\\[0.1133ex]
\label{ImU}
{\rm Im} [U(\vec r_A, \vec r_B, \vec R)] = & \;
\left[ \beta^{ij} \, \left( \sin( \omega_0 \, R ) -
\omega_0 R \, \cos( \omega_0 \, R ) \right) 
- \alpha^{ij} \, \omega_0^2 \, R^2 \, \sin( \omega_0 \, R ) 
\right] 
\, \frac{d_A^i \, d_B^j}{4 \pi R^3} \,.
\end{align}
\end{subequations}
{\color{black} The real part had previously been
given in Eq.~(5) of Ref.~\cite{MLPo1964}
and in Eq.~(7.2.27) on page 149 in Chapter 7
of Ref.~\cite{CrTh1984}.}
We have used the relations given in Eq.~\eqref{Dij}.
Using $\hat R = \vec R/R$, one has 
the static limit ($\omega_0 R \to 0$), 
\begin{equation}
U(\vec r_A, \vec r_B, \vec R) \to 
\left( \delta_{ij}- 3 \hat R_i \, \hat R_j  \right)
\frac{ d_{Ai} \, d_{Bj} }{4 \pi R^3} \, 
\qquad
\omega_0 \to 0 \,,
\end{equation}
which verifies the well-known expression for the
non-retarded van-der-Waals interaction 
given in Eq.~\eqref{HvdW}. SI units can be restored by 
multiplication with an additional overall factor $1/\epsilon_0$,
and replacing the factor $\omega_0 \, R$ by the 
(dimensionless) factor $\omega_0 \, R/c$.

\begin{figure}
\begin{center}
\begin{minipage}{0.8\linewidth}
\begin{center}
\includegraphics[width=0.4\linewidth]{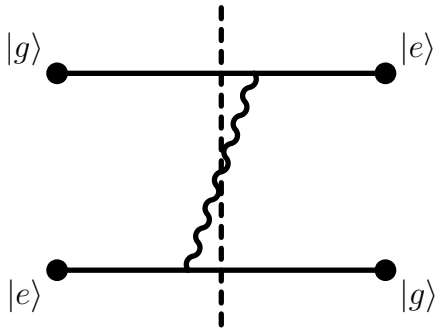}
\caption{\label{fig2} When the resonant virtual 
exchange photon becomes on-shell (a real photon),
a decay process is being described.
{\color{black}
By the Cutkosky rules~\cite{Cu1960}, this is denoted
by a vertical dashed line which cuts the diagram.
Only one of the diagrams in Fig.~\ref{fig1} 
contributes to the imaginary part.}
The internal state of the diagram,
which is cut open, has both atoms in the ground state and the
photon becoming real. Hence, it is 
a valid final state for the decay process
{\color{black} (see also Fig.~1 of Ref.~\cite{An1989},
Sec.~2.1 of Ref.~\cite{Sa2015} and Ref.~\cite{JeDaAn2004})}.
However, the sign of the imaginary part of 
the exchange interaction potential given
in Eq.~\eqref{ImU} is oscillating.
For a decay rate to be described by 
the imaginary part of an energy shift,
we would not expect such an oscillation;
hence, a careful interpretation is required.}
\end{center}
\end{minipage}
\end{center}
\end{figure}

%
% Interpretation of the Imaginary Part
%
\section{Interpretation of the Imaginary Part}
\label{sec4}

The imaginary part of the interatomic interaction
potential given in Eq.~\eqref{ImU} is oscillating in 
sign. It describes the one-photon resonant 
emission from the decaying excited state
which forms part of the entangled Dicke 
states given in Eq.~\eqref{dicke}.
in full analogy, the imaginary part of the 
one-loop self-energy of an excited bound 
state in atom is naturally interpreted 
in terms of the decay width of that same 
excited reference state~\cite{BaSu1978,Je2007,Je2008}.
The corresponding Feynman diagram is 
given in Fig.~\ref{fig2}.
The oscillating character of the imaginary part
could be seen as a problem because
the imaginary part of the 
resonance energy of a decaying state
is required to be negative~\cite{Mo1998,JeSuZJ2009prl};
a positive imaginary part would correspond
to an anti-resonance and a negative decay rate.
{\color{black} The solution is presented 
in the following. We can 
anticipate that the negative decay rate
which could otherwise naively result from the interaction
energy is compensated by the natural decay rate,
to give the subradiant and superradiant states their
(nonnegative) decay rates.}

One generally writes a resonance energy in terms of 
${\rm Re} \, E - \ii \Gamma/2$, where
$\Gamma$ is the width. Hence, it is useful
to define a decay rate operator
$\widehat{\Gamma}(\xi)$, where 
$\xi = \omega_0 \, R$ is the dimensionless 
argument appearing in Eq.~\eqref{ImU}.
This operator is 
related to the imaginary part of the 
exchange potential as follows,
\begin{align}
\widehat{\Gamma}(\xi) =& \;
-2 \; {\rm Im} [U(\vec r_A, \vec r_B, \vec R)]  \,,
\qquad
\xi = \omega_0 \, R \,.
\end{align}
We write the result given in Eq.~\eqref{ImU} 
somewhat differently as
\begin{align}
\widehat{\Gamma}(\xi) =& \; \frac32
\frac{ \xi \, \cos( \xi )
+ [ \xi^2 - 1 ]\, \sin( \xi ) }{\xi^3 } \,
\frac{ 4 \alpha }{ 3 } \, 
\omega_0^3 \, 
\left( \delta^{ij} - \frac{R^i \, R^j}{R^2} \right) 
r_A^i \, r_B^j
\nonumber\\
& \; +
3 \frac{ \sin( \xi ) - \xi \, \cos( \xi ) }{ \xi^3 }
\frac{4 \alpha}{3} \, \omega_0^3 \, 
\frac{R^i \, R^j}{R^2} \, r_A^i \, r_B^j 
\end{align}
This expression can be conveniently 
written in terms of a
transverse decay operator $\widehat{\Gamma}_\perp(\xi)$, which describes
transition whose polarization axis is perpendicular to the 
interatomic distance vector $\vec R$, and a 
longitudinal decay operator $\widehat{\Gamma}_\parallel(\xi)$, which describes
transition whose polarization axis is parallel to the 
interatomic distance vector $\vec R$.
The result is 
\begin{equation}
\widehat{\Gamma}(\xi) =
\widehat{\Gamma}_\perp(\xi) +
\widehat{\Gamma}_\parallel(\xi)  \,,
\qquad
\widehat{\Gamma}_\perp(\xi) =
f_\perp(\xi) \, \widehat{\Gamma}_\perp(0) \,,
\qquad
\widehat{\Gamma}_\parallel(\xi) =
f_\parallel(\xi) \, \widehat{\Gamma}_\parallel(0) \,.
\end{equation}
The weight functions $f_\perp(\xi)$ and 
$f_\parallel(\xi)$ and the 
transverse and longitudinal decay operator
at zero distance, $\widehat{\Gamma}_\perp(0)$ and
$\widehat{\Gamma}_\parallel(0)$, are given as follows,
\begin{align}
\widehat{\Gamma}_\perp(0) = & \;
\frac{ 4 \alpha }{ 3 } \, \omega_0^3 \,
\left( \delta^{ij} - \frac{R^i \, R^j}{R^2} \right)
r_A^i \, r_B^j \,,
\\
\widehat{\Gamma}_\parallel(0) = & \;
\frac{4 \alpha}{3} \, \omega_0^3 \,
\frac{R^i \, R^j}{R^2} \, r_A^i \, r_B^j \,,
\\
\widehat{\Gamma}(0) = & \; \widehat{\Gamma}_\perp(0) +
\widehat{\Gamma}_\parallel(0) = 
\frac{4 \alpha}{3} \, \omega_0^3 \, 
\delta^{ij} \, r_A^i \, r_B^j \,,
\\
\label{perp}
f_\perp(\xi) =& \; 3 \frac{ \xi \cos(\xi)
+ [ \xi^2 - 1] \, \sin( \xi )}{2 \xi^3} 
= 1 + \mathcal{O}(\xi^2) \,,
\\
\label{parallel}
f_\parallel(\xi) =& \;
3 \, \frac{ \sin(\xi) - \xi \, \cos(\xi) }{ \xi^3 }
= 1 + \mathcal{O}(\xi^2) \,,
\end{align}
Here, $\widehat{\Gamma}(0)$ is the total decay operator
at zero distance. It fulfills the relation
$\langle \psi_e \, \psi_g | \widehat{\Gamma}(0) | \psi_g \, \psi_e \rangle =
\Gamma(0)$, where $\Gamma(0)$ (no hat symbol over the $\Gamma$) 
is the natural decay width of the excited state in vacuum,
which is obtained without consideration of the 
atom-atom-interaction and its consequential 
modification of the decay rate.
One compares it with Eq.~(3.37) of Ref.~\cite{JeAd2022book}.

Let us now investigate the two Dicke states,
\begin{align}
| \Psi_\pm \rangle 
=& \; \frac{1}{\sqrt{2}} \,
\left[ | \psi_e \, \psi_g \rangle
\pm | \psi_g \, \psi_e \rangle \right] \,.
\end{align}
The expectation value of the decay rate operator
can be written as follows,
\begin{align}
\langle \Psi_\pm | \widehat{\Gamma}(\xi) | \Psi_\pm \rangle 
=& \; \pm \langle \psi_g \, \psi_e | \widehat{\Gamma}(\xi) | \psi_e \, \psi_g \rangle  
\\
=& \; \pm \left[ f_\perp(\xi) \, \Gamma^{eg}_\perp(0) 
+ f_\parallel(\xi) \, \Gamma^{eg}_\parallel(0) \right] \,.
\end{align}
Here,
\begin{align}
\Gamma^{eg}_\perp(0) = & \;
\langle \psi_g \, \psi_e | \widehat{\Gamma}_\perp(0) | \psi_e \, \psi_g \rangle  
= \frac{ 4 \alpha }{ 3 } \, \omega_0^3 \,
\left( \delta^{ij} - \frac{R^i \, R^j}{R^2} \right)
\langle g | r_A^i | e \rangle \, 
\langle e | r_B^j | g \rangle 
\\
=& \; \frac{ 4 \alpha }{ 3 } \, \omega_0^3 \,
\left( \delta^{ij} - \frac{R^i \, R^j}{R^2} \right)
\langle g | r_A^i | e \rangle \,
\langle e | r_A^j | g \rangle \,,
\\
\Gamma^{eg}_\parallel(0) = & \;
\langle \psi_g \, \psi_e | \widehat{\Gamma}_\parallel(0) | \psi_e \, \psi_g \rangle
= \frac{ 4 \alpha }{ 3 } \, \omega_0^3 \,
\frac{R^i \, R^j}{R^2} \,
\langle g | r_A^i | e \rangle \,
\langle e | r_B^j | g \rangle
\\
=& \; 
\frac{ 4 \alpha }{ 3 } \, \omega_0^3 \,
\frac{R^i \, R^j}{R^2} \,
\langle g | r_A^i | e \rangle \,
\langle e | r_A^j | g \rangle \,,
\\
\Gamma(0) =& \;
\Gamma^{eg}_\perp(0) + \Gamma^{eg}_\parallel(0) = 
\frac{ 4 \alpha }{ 3 } \, \omega_0^3 \,
| \langle g | \vec r_A | e \rangle |^2 \,.
\end{align}
We have repeatedly used the fact that the 
two atoms are identical, to replace $r_B^j \to r_A^j$.
Note that $\Gamma^{eg}_\perp(0)$ and 
$\Gamma^{eg}_\parallel(0)$ can depend on the 
magnetic projections,
but the natural width $\Gamma(0)$ is independent 
of magnetic quantum numbers~\cite{JeAd2022book}.

In the space spanned by the 
$ | \psi_g \psi_e \rangle$ and 
$ | \psi_e \psi_g \rangle$, one finally obtains
the following effective Hamiltonian
for the two-atoms system,
which takes into account the one-photon exchange
and, with it, the imaginary part of the 
exchange energy,
\begin{multline}
\mathbbm{H} = 
\left[ E_g + E_e - \tfrac{\ii}{2} \Gamma(0) \right] \, 
\left( | \psi_g \psi_e \rangle \, \langle \psi_g \psi_e | 
+ | \psi_e \psi_g \rangle \, \langle \psi_e \psi_g | 
\right) 
\\
+ \left\{ | \psi_g \, \psi_e \rangle \, 
\left\{ {\rm Re} \, U(\xi) -\frac{\ii}{2}
\left[ f_\perp(\xi) \, \Gamma_\perp(0) 
+ f_\parallel(\xi) \, \Gamma_\parallel(0) \right]
\right\}
\langle \psi_e \, \psi_e |  + 
\mbox{h.c.} \right\}
\end{multline}
Here, $E_e = {\rm Re} E_e$ is the real part of the 
energy of the excited state,
and we have supplemented the term
$-\ii \Gamma(0)/2$ in the resonance energy of the 
unperturbed states.
It means that in the basis of states
$| \psi_g \psi_e \rangle$,
$| \psi_e \psi_g \rangle$, the Hamiltonian matrix is
\begin{equation}
\mathbbm{H} = \left( \begin{array}{cc}
E_0 & \delta E \\
\delta E & E_0 \\
\end{array} \right) \,,
\qquad 
E_0 =  E_g + E_e - \frac{\ii}{2} \, \Gamma(0) \,,
\qquad 
\delta E = E_\gamma - \frac{\ii}{2} \, \Gamma_\gamma \,,
\end{equation}
Here, 
\begin{equation}
E_\gamma = \langle \psi_g \, \psi_e | 
{\rm Re} \, U(\xi) | \psi_g \, \psi_e \rangle \,,
\qquad
\Gamma_\gamma = \langle \psi_g \, \psi_e | 
\widehat{\Gamma}(\xi) | \psi_g \, \psi_e \rangle \,,
\end{equation}
describe the real and imaginary parts of the one-photon exchange energy.
We write the vector representation of the Dicke states 
$| \Psi_- \rangle$ and $| \Psi_+ \rangle$ and
$E_- = E_-(\xi)$, and $E_+ = E_+(\xi)$, as follows,
\begin{equation}
| \Psi_\pm \rangle = \frac{1}{\sqrt{2}} \,
\left( \begin{array}{c} 1 \\ \pm 1 \end{array} \right) \,,
\qquad
E_\pm = E_0 \pm \delta E \,.
\end{equation}
The resonance energies can be written as follows,
\begin{align}
E_-(R) =& \; {\rm Re} \, E_0 + E_\gamma - \frac{\ii}{2} \left\{ 
\Gamma(0) 
- \left[ f_\perp(\xi) \, \Gamma^{eg}_\perp(0) 
+ f_\parallel(\xi) \, \Gamma^{eg}_\parallel(0) \right] \right\} \,,
\\
E_+(R) =& \; {\rm Re} \, E_0 - E_\gamma - \frac{\ii}{2} \left\{ \Gamma(0) + 
\left[ f_\perp(\xi) \, \Gamma^{eg}_\perp(0) 
+ f_\parallel(\xi) \, \Gamma^{eg}_\parallel(0) \right] \right\} \,.
\end{align}
The effective decay rates of the Dicke states are
\begin{align}
\label{Gamma_minus}
\Gamma_-(R) =& \; \Gamma(0) 
- \left[ f_\perp(\xi) \, \Gamma^{eg}_\perp(0) 
+ f_\parallel(\xi) \, \Gamma^{eg}_\parallel(0) \right] = 
\mathcal{O}(\xi^2) \,,
\\
\label{Gamma_plus}
\Gamma_+(R) =& \; \Gamma(0) +
\left[ f_\perp(\xi) \, \Gamma^{eg}_\perp(0) 
+ f_\parallel(\xi) \, \Gamma^{eg}_\parallel(0) \right] 
= 2 \Gamma(0) - \mathcal{O}(\xi^2) \,,
\end{align}
Hence, 
$| \Psi_- \rangle$ is the entangled subradiant state,
while $| \Psi_+ \rangle$ is the superradiant state.
In the short-range limit, the subradiant state
$| \Psi_- \rangle$ becomes metastable in view of the 
entanglement, while the superradiant state
$| \Psi_+ \rangle$ acquires twice the 
natural decay width.

%
% Verification of the Result
%
\section{Verification of the Result}
\label{sec5}

Let us include some cryptic remarks regarding 
a verification of the results given in Eqs.~\eqref{Gamma_minus} 
and~\eqref{Gamma_plus} based on an alternative method.
To this end, one considers Eq.~(7.2.4) of Ref.~\cite{CrTh1984},
but crucially takes into account that the 
entangled photon emissions from the atoms $A$ and $B$ 
happen at different positions.
Thus, one considers the matrix element $M$,
\begin{align}
\label{M}
M =& \; \tfrac12 \left| \left< \psi_g \, \psi_g \left| 
\hat \epsilon \cdot \vec r_A \; \ee^{\ii \vec k \cdot \vec R_A} +
\hat \epsilon \cdot \vec r_B \; \ee^{\ii \vec k \cdot \vec R_B}
\right| \psi_e \, \psi_g \pm \psi_g \, \psi_e \right> \right|^2
\nonumber\\
=& \;
\left| \left< \psi_g \left|
\hat \epsilon \cdot \vec r_A
\right| \psi_e \right> \right|^2
\pm 
\left| \left< \psi_g \left| \hat \epsilon \cdot \vec r_A \right| \psi_e \right>
\right|^2 \, \cos( \vec k \cdot \vec R )  \,.
\end{align}
where $\vec R = \vec R_A - \vec R_B$.
In the last step, we have used the indistinguishability 
of the two identical atoms.
Furthermore, on resonance, one has $| \vec k | = \omega_0$.
The sum over the two photon polarizations results in
\begin{equation}
\sum_\lambda \epsilon_\lambda^i \, \epsilon_\lambda^j
= \delta^{ij} - \frac{k^i \, k^j}{\vec k^2} \,,
\end{equation}
After some algebra, one shows the result
\begin{align}
\label{Omega}
\int \dd \Omega \frac{3}{8 \pi}
\left( \delta^{ij} - \frac{k^i \, k^j}{\vec k^2} \right) \,
\cos( \omega_0 \hat k \cdot \vec R ) =
\left( \delta_{ij} - \frac{R^i \, R^j}{\vec R^2} \right) \,
\, f_\perp(\omega_0 R) +
\frac{R^i \, R^j}{\vec R^2} f_\parallel(\omega_0 R) \,,
\end{align}
where the previously obtained weight functions
$f_\perp(\omega_0 R)$ and 
$f_\parallel(\omega_0 R)$ [see Eqs.~\eqref{perp} and~\eqref{parallel}]
naturally appear.
Based on these identities and the formalism outlined
in Chaps.~3 and 4~of Ref.~\cite{JeAd2022book}, 
one can verify the results given in Eqs.~\eqref{Gamma_minus}
and~\eqref{Gamma_plus}.
The first term on the right-hand side of Eq.~\eqref{Omega}
gives the term $\Gamma(0)$ in 
Eqs.~\eqref{Gamma_minus} and~\eqref{Gamma_plus}, 
while the {\color{black} second} term on the right-hand side of Eq.~\eqref{Omega}
gives the second term on the right-hand side of 
Eqs.~\eqref{Gamma_minus} and~\eqref{Gamma_plus}.

Conversely, the introduction of the Feynman prescription
into the formalism outlined in Ref.~\cite{CrTh1984},
which was initially designed to obtain the real 
part of the resonant one-photon exchange interaction only,
leads to the imaginary part obtained here.
Notes on this point are collected in the Appendix.

%
% Example of Superradiant and Subradiant States
%
\section{Example of Superradiant and Subradiant States}
\label{sec6}

We would like to conclude this work with a concrete example,
namely, the hydrogen atom, where the $1S$ state is the ground state
and the excited state could be one of three $2P$ states,
which we take in a Cartesian basis.
Furthermore, we assume that the interatomic 
distance is along the $z$ axis,
\begin{equation}
\vec R = R \, \hate_z \,.
\end{equation}
The wave functions are well known,
\begin{subequations}
\begin{align}
\psi_g(\vec r) =& \; \psi_{1S}(\vec r) = 
\frac{1}{\sqrt{\pi} \, a_0^{3/2}} \,
\exp\left(-\frac{r}{a_0} \right) \,,
\qquad
r = |\vec r| \,,
\\[0.1133ex]
\psi_{e,x}(\vec r) =& \; \psi_{2P,x}(\vec r) = 
\frac{r \, \sin \theta \, \cos\varphi}{4 \sqrt{2\pi} \, a_0^{5/2} } \,
\exp\left(-\frac{r}{2 a_0} \right) \,,
\\[0.1133ex]
\psi_{e,y}(\vec r) =& \; \psi_{2P,y}(\vec r) = 
\frac{r \, \sin \theta \, \sin\varphi}{4 \sqrt{2\pi} \, a_0^{5/2} } \,
\exp\left(-\frac{r}{2 a_0} \right) \,,
\\[0.1133ex]
\psi_{e,z}(\vec r) =& \; \psi_{2P,z}(\vec r) = 
\frac{r \, \cos \theta }{4 \sqrt{2\pi} \, a_0^{5/2} } \,
\exp\left(-\frac{r}{2 a_0} \right) \,.
\end{align}
\end{subequations}
The nonvanishing dipole transition matrix elements are as follows,
\begin{equation}
\langle \psi_g | x | \psi_{e,x} \rangle = 
\langle \psi_g | y | \psi_{e,y} \rangle = 
\langle \psi_g | z | \psi_{e,z} \rangle = 
\frac{2^7}{3^5} \, \sqrt{2} \, a_0 \,,
\end{equation}
while all other combinations vanish
(e.g., $\langle \psi_g | x | \psi_{e,y} \rangle = 0$).
The resonance frequency is
\begin{equation}
\omega_0 = \frac38 \, \alpha^2 m
\end{equation}
and the natural decay width is
(see Ref.~\cite{JeMo2002})
\begin{equation}
\Gamma(0) = \frac{2^8}{3^8} \alpha^5 m \,.
\end{equation}
The $x$ and $y$ polarized $P$ states
cannot decay via a $z$-polarized transition,
while the $z$ polarized $P$ states
decays exclusively via a $z$-polarized transition.
We have the results
\begin{equation}
\Gamma_\perp^{0x}(0) = \Gamma_\perp^{0y}(0) = 
\Gamma_\parallel^{0z}(0) = \Gamma(0) \,,
\qquad
\Gamma_\parallel^{0x}(0) = \Gamma_\parallel^{0y}(0) = 0 \,,
\qquad
\Gamma_\perp^{0z}(0) = 0 \,.
\end{equation}
One can thus define the distance-dependent 
transverse and longitudinal decay rates,
$\Gamma_\perp(\xi)$ and $\Gamma_\parallel(\xi)$,
\begin{align}
\Gamma_\perp(\xi) =& \; 
\Gamma_\perp^{0x}(\xi) = \Gamma_\perp^{0y}(\xi) =
f_\perp(\xi) \, \Gamma(0) \,,
\\
\Gamma_\parallel(\xi) =& \; 
\Gamma^{0z}_\parallel(\xi) =
f_\parallel(\xi) \, \Gamma(0)  \,,
\qquad
\xi = \omega_0 \, R \,.
\end{align}
At short range, the decay-rate admixtures expand as follows,
\begin{align}
\frac{ \Gamma_\perp(R) }{ \Gamma(0) } =& \;
1 - \frac{(\omega_0 R)^2}{5} + \mathcal{O}((\omega_0 R)^4) \,,
\\
\frac{ \Gamma_\parallel(R) }{ \Gamma(0) } =& \;
1 - \frac{(\omega_0 R)^2}{10} + \mathcal{O}((\omega_0 R)^4) \,,
\end{align}

\begin{figure}[t!]
\begin{center}
\begin{minipage}{0.8\linewidth}
\begin{center}
\includegraphics[width=0.8\linewidth]{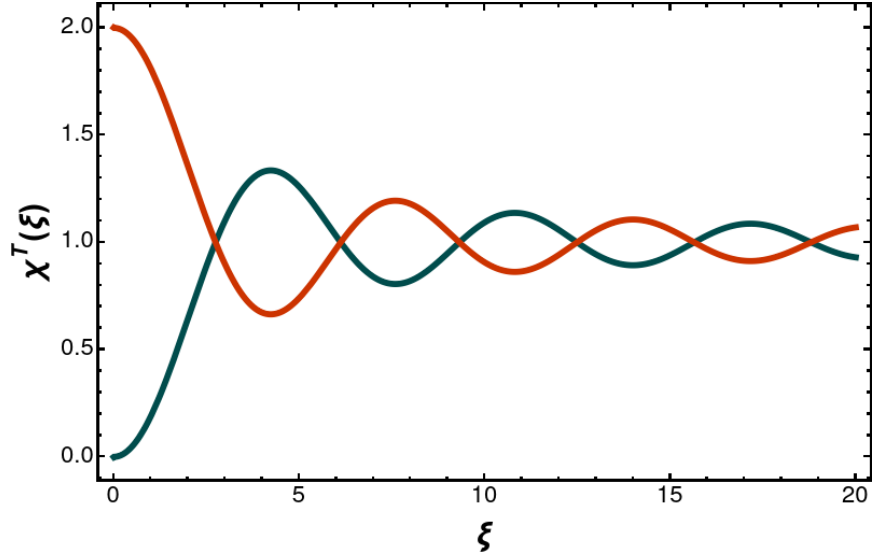}
\caption{\label{fig3} The plot shows the 
function $\chi^{\rm T}(\xi) = 1 \pm f_\perp(\xi)$, 
which is the ratio of the effective decay rate
of the superradiant and subradiant entangled Dicke
states as a function of the interatomic distance.
The red curve corresponds to the positive 
sign (superradiant), while the blue curve corresponds to the 
negative sign (subradiant state). 
The decay rate of the transverse superradiant state, in the short-range
limit, assumes a value equal to twice the natural decay rate.}
\end{center}
\end{minipage}
\end{center}
\end{figure}

From the $x$ or $y$ polarized $P$ states,
we have the following subradiant entangled Dicke states, 
\begin{equation}
| \Psi^x_- \rangle = \frac{1}{\sqrt{2}} \, ( | \psi_g \, \psi_{e,x} \rangle -
| \psi_{e,x} \, \psi_g \rangle )\,,
\qquad
| \Psi^y_- \rangle = \frac{1}{\sqrt{2}} \, ( | \psi_g \, \psi_{e,y} \rangle -
| \psi_{e,y} \, \psi_g \rangle )\,.
\end{equation}
They have the following decay rates,
\begin{align}
\Gamma^\perp_-(R) =& \; \Gamma(0) - \Gamma_\perp(R)
= \Gamma(0) \, [ 1 - f_\perp(\xi) ]
\\
=& \; \Gamma(0) \times 
\left\{ \begin{array}{cc}
\left[ \frac{(\omega_0 R)^2}{5} + \mathcal{O}((\omega_0 R)^4) \right] 
& \omega_0 \, R \to 0 \\[2.1133ex]
\left[ 1 - \frac{3 \sin(\omega_0 R)}{2 \omega_0 R} 
+ \mathcal{O}\left( \frac{1}{(\omega_0 R)^2} \right) \right] 
& \omega_0 \, R \to \infty \\ \end{array} \right. \,.
\end{align}
In the short-range limit, the decay is suppressed,
while in the long-range limit, the natural decay width is
approached, albeit with a long-range, sinusoidal modification. 
For the $z$ polarized (longitudinal), subradiant state, 
\begin{equation}
| \Psi^\parallel_- \rangle = \frac{1}{\sqrt{2}} \, ( | \psi_g \, \psi_{e,z} \rangle -
| \psi_{e,z} \, \psi_g \rangle )\,,
\end{equation}
one finds for the decay rate
\begin{align}
\Gamma^\parallel_-(R) =& \; \Gamma(0) - \Gamma_\parallel(R)
= \Gamma(0) \, [ 1 - f_\parallel(\xi) ]
\\[3ex]
=& \; \Gamma(0) \times 
\left\{ \begin{array}{cc}
\left[ \frac{(\omega_0 R)^2}{10} + \mathcal{O}( (\omega_0 R)^4) \right]
& \omega_0 \, R \to 0 \\[2.1133ex]
\left[ 1 + \frac{3 \cos(\omega_0 R)}{(\omega_0 R)^2}
+ \mathcal{O}\left( \frac{1}{(\omega_0 R)^4} \right) \right]
& \omega_0 \, R \to \infty \\ \end{array} \right. \,.
\end{align}

\begin{figure}[t!]
\begin{center}
\begin{minipage}{0.8\linewidth}
\begin{center}
\includegraphics[width=0.8\linewidth]{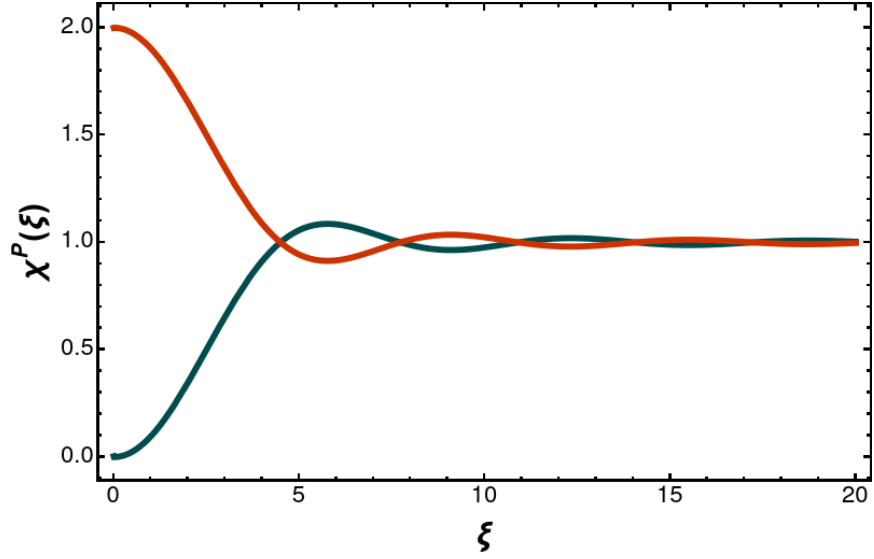}
\caption{\label{fig4} The plot shows the 
function $\chi^{\rm P}(\xi) = 1 \pm f_\parallel(\xi)$, 
where the red curve corresponds to the positive 
sign, while the blue curve corresponds to the 
negative sign. The positive sign corresponds to the longitudinal
superradiant state, whose decay rate, in the short-range
limit, assumes a value equal to twice the natural decay rate.}
\end{center}
\end{minipage}
\end{center}
\end{figure}

From the $x$ or $y$ polarized $P$ states,
we calculate the following superradiant entangled Dicke states,
\begin{equation}
| \Psi^x_+ \rangle = \frac{1}{\sqrt{2}} \, ( | \psi_g \, \psi_{e,x} \rangle +
| \psi_{e,x} \, \psi_g \rangle )\,,
\qquad
| \Psi^y_+ \rangle = \frac{1}{\sqrt{2}} \, ( | \psi_g \, \psi_{e,y} \rangle +
| \psi_{e,y} \, \psi_g \rangle )\,.
\end{equation}
One finds the decay rates,
\begin{align}
\Gamma^\perp_+(R) =& \; 
\Gamma(0) + \Gamma_\perp(R) = 
\Gamma(0) \, [ 1 + f_\perp(\xi) ]
\\
=& \; \Gamma(0) \times
\left\{ \begin{array}{cc}
\left[ 2 - \frac{(\omega_0 R)^2}{5} + \mathcal{O}(R^4) \right]
& \omega_0 \, R \to 0 \\[2.1133ex]
\left[ 1 + \frac{3 \sin(\omega_0 R)}{2 \omega_0 R}
+ \mathcal{O}\left( \frac{1}{(\omega_0 R)^2} \right) \right]
& \omega_0 \, R \to \infty \\ \end{array} \right. \,,
\end{align}
In the short-range limit, the decay rate is twice the 
natural width, while, in the long-range limit, 
the natural width is approached, with a long-range
sinosoidal modification.
One also finds the following superradiant longitudinal 
entangled Dicke state,
\begin{equation}
| \Psi^\parallel_+ \rangle = \frac{1}{\sqrt{2}} \, ( | \psi_g \, \psi_{e,z} \rangle +
| \psi_{e,z} \, \psi_g \rangle )\,.
\end{equation}
The distance-dependent decay rate is
\begin{align}
\Gamma^\parallel_+(R) =& \; \Gamma(0) + \Gamma_\parallel(R)
\Gamma(0) \, \chi_+^\parallel(\omega_0 R) 
\\
=& \; \Gamma(0) \times
\left\{ \begin{array}{cc}
\left[ 2 - \frac{(\omega_0 R)^2}{10} + \mathcal{O}((\omega_0 R)^4) \right]
& \omega_0 \, R \to 0 \\[2.1133ex]
\left[ 1 - \frac{3 \cos(\omega_0 R)}{ ( \omega_0 R )^2 }
+ \mathcal{O}\left( \frac{1}{(\omega_0 R)^3} \right) \right]
& \omega_0 \, R \to \infty \\ \end{array} \right. \,.
\end{align}
The results are illustrated in Figs.~\ref{fig3}
and~\ref{fig4}.

%
% Conclusions
%
\section{Conclusions}
\label{sec7}

We have considered retardation corrections to the 
one-photon exchange between identical atoms.
In Sec.~\ref{sec2}, we have considered the 
matching of the $S$ matrix element and the 
effective Hamiltonian.
Specifically, our calculation 
has been based on the non-forward scattering matrix 
element induced by an interaction potential
of the functional form $U(\vec r_A, \vec r_B, \vec R)$
which depends on the relative coordinates $\vec r_A$
and $\vec r_B$ of the two atoms,
and the interatomic distance $\vec R$.
For the matching to be successful, we need to 
assume the initial and final states of the process
to have identical energy. 
This is the case if, e.g., the initial state
is a combination of one of the atoms in the 
ground state, and the other atom in an 
excited $P$ state. The final state has the 
energies of the two states reversed,
with the possibility of different 
degeneracy indices for the initial and final
states. Because the energies of the quantum states of the individual 
states have been exchanged
in the initial and final states of the identical 
atoms, the total energy of the 
final state is equal to that of the initial state,
and the matching of the non-forward $S$ matrix 
element to the effective Hamiltonian can proceed.

This program is realized in Sec.~\ref{sec3},
where the calculation is realized in the 
temporal gauge for the virtual photon. In this gauge,
the timelike component of the photon 
propagator vanishes, which implies that it is the ideal 
gauge for the calculation of the retardation corrections
to the van der Waals potential given in Eq.~\eqref{HvdW}.
The final result is given in Eq.~\eqref{U}.
The retarded potential has both a real and an 
imaginary part.  The imaginary part is suppressed 
for the short-range interaction, as an 
expansion of Eq.~\eqref{ImU} for $R \ll a_0/\alpha$
immediately shows.
The interpretation of the imaginary part of the 
resonant one-photon exchange energy is 
discussed in Sec.~\ref{sec4}. In the quantum-field
theoretical picture,
the imaginary part is connected with the 
virtual resonant exchange photon becoming on shell,
and thus, being emitted as a real photon by the 
entangled two-atom system.
It is found that a completely consistent 
picture is obtained when one calculates
the Hamiltonian matrix including the 
unperturbed resonance energies of the 
ground and excited states, as well as the 
one-photon resonant exchange energy and its
imaginary part. The modification of the 
decay rate of the entangled Dicke state is consistently 
obtained, and the result allows us to obtain
consistent formulas for the distance-dependent 
decay rates of the superradiant and subradiant Dicke states.

A possibility for an independent verification 
of the results based on Fermi's Golden Rule is sketched 
in Sec.~\ref{sec5}. An example calculation 
involving hydrogen $1S$ and $2P$ states is given in 
Sec.~\ref{sec6}, culminating in the results
presented in Fig.~\ref{fig3} and~\ref{fig4}.
The modification of the decay rates has a
long-range, $1/R$ tail. Our calculations show that the
imaginary part of the retarded one-photon exchange 
as obtained by the Feynman prescription finds 
a completely consistent physical interpretation in terms
of the distance-dependent modification of the 
decay rate of entangled superradiant and subradiant Dicke states
of the identical two-atom system.

%
% Example of a Hydrogen Atom
%
\section{Appendix: Comparison with the Literature}
\label{app}

In Chap.~7 of Ref.~\cite{CrTh1984}, the authors
carry out a related calculation using time-ordered 
perturbation theory. 
Specifically, in Eq.~(7.2.21) of Ref.~\cite{CrTh1984}, 
the authors obtain an expression for the retarded 
van der Waals interaction which, in our notation, 
reads as follows,
\begin{align}
\label{23}
U'(\vec r_A, \vec r_B, \vec R)] = & \;
\lim_{\gamma \to 0^+} 
\frac{1}{2 \pi^2} \int_0^\infty \dd p \, p^4 \, 
\frac{\tau_{ij}(p \, R)}{\omega_0^2 - p^2} 
\, \ee^{-\gamma p} \,,
\\[0.1133ex]
\label{taudef}
\tau_{ij}(p \, R) =& \;
\alpha_{ij} \frac{\sin(p \, R)}{p \, R} 
+ \beta_{ij} \left( \frac{\cos(p \, R)}{p^2 \, R^2} -
\frac{\sin(p \, R)}{p^3 R^3} \right) \,.
\end{align}
For clarity, we observe 
that $\omega_0$ is denoted as $k$ in Ref.~\cite{CrTh1984}, 
and that $\epsilon_0$ is explicitly written 
out in Ref.~\cite{CrTh1984}.
In Chap.~7 of Ref.~\cite{CrTh1984}, the authors evaluate 
the integral over $p$ as a principal-value integral,
using a convergent factor, and obtain the 
real part of our result,
\begin{equation}
\label{conf}
U'(\vec r_A, \vec r_B, \vec R)] = 
{\rm Re} [U(\vec r_A, \vec r_B, \vec R)] \,.
\end{equation}
The Feynman prescription of quantum electrodynamics is 
implemented by the substitution
\begin{equation}
\frac{\tau_{ij}(p \, R)}{\omega_0^2 - p^2} \to
\frac{\tau_{ij}(p \, R)}{\omega_0^2 - p^2 + \ii \epsilon} 
= \frac{\tau_{ij}(p \, R)}{(\omega_0 +\ii \epsilon - p) \,
(\omega_0 +\ii \epsilon + p) }
\end{equation}
in Eq.~\eqref{23}. As a function of $p$, the integrand
then obtains poles in the complex $p$ plane at 
$p = \pm \sqrt{\omega_0^2 + \ii \epsilon}$,
i.e., at $p = \omega_0 + \ii \epsilon$ and at
$p = -\omega_0 - \ii \epsilon$,
where $\epsilon$ denotes an infinitesimal 
positive imaginary part.
In order to consider the contribution from the 
pole, one symmetrizes the integral on the 
domain $-\infty < p < \infty$, and inserts the 
infinitesimal imaginary part,
leading to 
\begin{equation}
\label{Upp}
U''(\vec r_A, \vec r_B, \vec R)] = 
\lim_{\epsilon\to 0^+} \lim_{\gamma \to 0^+}
\frac{1}{4 \pi^2} \int_{-\infty}^\infty \dd p \, p^4 \, 
\frac{\tau_{ij}(p \, R)}{\omega_0^2 - p^2 + \ii \epsilon}
\, \ee^{-\gamma |p|} \,.
\end{equation}
The real part of 
$U''(\vec r_A, \vec r_B, \vec R)$ can be evaluated
by principal value,
confirming the result given in Eq.~\eqref{conf}.
The imaginary part is obtained by considering the 
pole at $p = \omega_0 + \ii \epsilon$ which is 
encircled in the mathematically positive sense,
\begin{equation}
\ii \, {\rm Im} U''(\vec r_A, \vec r_B, \vec R)] = 
2 \pi \ii \, \mathop{\rm Res}_{p = \omega_0} 
\frac{1}{4 \pi^2} p^4 \, 
\frac{\tau_{ij}(p \, R)}{\omega_0^2 - p^2}
= -\frac{\ii}{4 \pi} \, \omega_0^3 \, \tau_{ij}(\omega_0 \, R)
= \ii \, {\rm Im} U(\vec r_A, \vec r_B, \vec R)] \,,
\end{equation}
where the latter equality is obvious by inspection 
of Eqs.~\eqref{taudef},~\eqref{Upp} and~\eqref{ImU}.
We conclude that,
if the infinitesimal imaginary part of the photon 
is supplemented in the treatment outlined in 
Chap.~7 of Ref.~\cite{CrTh1984}, then 
{\color{black} the result for the 
interaction potential $U$ given in Eq.~\eqref{U}
can be obtained,}
\begin{equation}
U''(\vec r_A, \vec r_B, \vec R) = U(\vec r_A, \vec r_B, \vec R) \,.
\end{equation}
We should also briefly discuss certain approximations
underlying our treatment and that discussed in 
Chap.~7 of Ref.~\cite{CrTh1984}.
First, we note the Born--Oppenheimer approximation,
which allows us to consider the motion of the 
electrons and nuclei separately and 
corresponds to the limit of infinitely heavy atomic 
nuclei (vanishing electron-nucleon mass ratio).
In the same light, we neglect the recoil energy 
upon photon emission, and we assume that the 
wave functions of the two atoms do not overlap.
Higher-order multipoles as well as multi-photon 
transitions are also neglected.
The same approximations underlie the treatments
given in Refs.~\cite{Va1970,Va1972} for 
systems consisting of unlike atoms with close 
excitation energies.

\acknowledgments{The authors acknowledge support by the National 
Science Foundation (grant PHY--2110294).}

\end{document}